\documentclass[twocolumn,pre,10pt,amsmath,amssymb,nofootinbib,showpacs,superscriptaddress,floatfix]{revtex4-2}
\usepackage{amsmath, mathrsfs, latexsym, amssymb, color, graphicx, comment, hyperref, scrhack}
\usepackage{lmodern}
\usepackage{physics}
\usepackage{booktabs}

\newcommand{\dfr}[2]{\frac {\displaystyle #1}{\displaystyle #2}}

\usepackage{multirow}
\usepackage{flushend}
\usepackage{fancyhdr}
\usepackage{todonotes}
\usepackage{graphicx}
\usepackage{color}

\newcommand{\e}[1] {Eq.~(\ref{#1})}

\begin{document}
\title{Competition of glass and crystal: phase-field model}

\author{M.\,G.\,Vasin}
\affiliation{Theoretical department, Vereshchagin Institute of High Pressure Physics, Russian Academy of Sciences, 108840 Moscow, Russia}
\author{V.\,Ankudinov}
\affiliation{Theoretical department, Vereshchagin Institute of High Pressure Physics, Russian Academy of Sciences, 108840 Moscow, Russia}
\affiliation{Udmurt Federal Research Center, Ural Branch of the Russian Academy of Science, 426000 Izhevsk, Russia}






\begin{abstract}
The phase-field model for the description of the solidification processes with the glass–crystal competition is suggested. The model combines the first-order phase transition  model in the phase-field formalism and gauge-field theory of glass transition. We present a self-consistent system of stochastic  motion equations for unconserved  order parameters describing the crystal-like short-range ordering and vitrification. It is shown, that the model qualitatively describes the glass–crystal competition during quenching with finite cooling speed. The nucleation of the crystalline phase at slow cooling speeds and low undercoolings proceeds by a fluctuation mechanism. The model demonstrates the tendency to amorphization with the increase of its cooling rate. 
\end{abstract}


\flushbottom
\maketitle
\thispagestyle{empty}

\section{Introduction}

The study and description of fundamental physical processes occurring during solidification remains an important task of the modern condensed matter physics \cite{Ojovan2008,Sperl2010,Xu2010,Tanaka2010}.
Among the current unsolved problems such as formation of the metastable structures on rapid propagating phase boundaries~\cite{gj19}, one can distinguish the rapid solidification of non-equilibrium melts with the competition between the amorphous and crystalline phases~\cite{Sanditov2019,Tropin2016,Berthier2011}. Modern experimental   setups for the metallic alloys processing can provide relatively high  undercoolings, temperature and concentration gradients. This techniques results in the rapid moving boundaries and huge interface velocities during phase transformations~\cite{herlach07}. Under these non-equilibrium conditions, the melt can solidify to the amorphous phase.
However, the problem of the theoretical description of  such transformations still remains unsolved.
This problem is related to a lack of understanding of physics of a glass transition which can be threated as a thermodynamic kinetically controlled phase transition \cite{Sanditov2019,Tropin2016,Berthier2011} or a topological transition~\cite{A7,A8,Jackle1986}. 

The objective of this work is to continuously extend a developed earlier theory of the competitive glass transition as a combination of the continous order parameter field and a topological order field \cite{Vasin2021,Vasin2022}.  Here we consider a  model with a concurence between two processes of ordering during cooling from a disordered liquid state. 

\subsection{Crystallization models (Phase-Field approach)}

We consider a phase transition from the liquid  to the crystalline phase as a transition with the order parameter responsible for the short-range ordering. The same idea is lead in the phase-field approach~\cite{bib:pe} where the phase state is determined by the order parameter variable with constant values related to the certain dedicated states. The phase-field (PF) and phase-field crystal (PFC) methodology are widely utilized in the materials science~\cite{bib:pe,ingstein,Emmerich2003,Boettinger2002}; such models are the generalization of the mean-field approximation in the theory of phase transitions. With the non-equilibrium thermodynamics approach~\cite{Jou2010,gj19}  PF and PFC can be  implemented for the description of the first- and second-order transitions in solidifying systems~\cite{Ankudinov20202}.  In the PF approach the state of the closed system is defined with the specific free energy which includes: (i) linear terms correspondent to the order parameter series expansion in certain form (typically double-well potential); (ii) non-linear terms correspondent to the phase interface or flux contributions in case of non-Markovian processes \cite{jg13,Jou2018,book,dlg14}.
The general idea of the modeling of dynamics with PF method is a requirement for a free energy decrease in time following from the Lyapunov condition~\cite{Jou2010}.
Dynamical equations may be formulated for conserved or non-conserved order parameters as well as mixed cases~\cite{bib:pe,ingstein,book} from the explicitly defined free energy. Phase-field models based on such thermodynamic approach are suitable to link the micro- and mesoscopic spatial scales as well as Brownian jump times and relatively long times of structural changes~\cite{ankudinov16,Ankudinov2020c,bib:pe,book}.

The application of such models to the non-ergodic amorphous phase states is not entirely clarified  yet since it requires going beyond the mean-field approximation and demands a description of the separate ordering field correspondent to the glassy state. In particular, several attempts to describe the transition to a glass-like state were made in \cite{Bruna2007,Galenko2019rsta,Galenko2020} with PF-model for single-component and binary systems; in these works the glass-forming phase is considered as a bulk phase with a slowed down diffusion strongly dependent on temperature. However, this model did not take into account a frustration, and non-ergodic effects inherent in glass, which are necessary to describe the formation of a fine-grained glass-like structures at high cooling rates.
In a number of papers, disordered amorphous phases are considered as analogue of liquids, when
crystallization (phase transition) is analogous to the transition  from liquid or amorphous phase and corresponds to the first-order phase transition~\cite{Dudorov2022,Gamov2012}.
 Study of glassy states in PFC encounters the same problems as with the molecular dynamics simulations such as challenging to the proper phase detection, long relaxation times and interference with noise-induced patterns~\cite{Toth2010,conti2016,Tang2014,Berry2014,Ankudinov2020c,Burns2022}.  The structural disorder can be observed in the  PFC, where the fast moving boundaries lead to the disorder trapping from the front and leading to the disordered glassy bulk \cite{Archer2012,Berry2011,Berry2008a}. One can observe the subsequenced delayed crystallization (and recrystallization) from amorphous nuclei to the crystalline bulk~\cite{Abdalla2022,Tang2017}.

There are some works, which reproduce experimental data of primary crystallization from the liquid phase and subsequent glass transition with PF~\cite{Rafique2018,Wu2015a,Wu2017}. In these studies, the glass transition is considered as a full-fledged phase transition, and the parameters of glassy state is considered in a bulk using the thermodynamical data obtained experimentally or with several approximations. Extended model \cite{Ericsson2019} introduces the phase-interface slowdown in a manner similar to~\cite{Bruna2007,Galenko2019rsta,Galenko2020}. To describe a glass state by itelf, a hybrid model consisting of a combination of a continuous mid-field model and a multiphase PF-model was developed \cite{Wang2012,Ericsson2019}. In the proposed model, the liquid-glass transition was considered with the noise-induced structural relaxation. This is a simplified model  that reproduces results comparable to the experiment. However, as well as other models to the authors' current knowledge  does not take into account the non-ergodicity and frustration. One can supply the possessed challenges with some additional problems emerged from the glass-liquid-solid interfaces description~\cite{Ganorkar2020,Ericsson2019}, confined amorphous states \cite{Cammarota2013}, and from the presence of the  long time relaxations leading to a nonlinear mobility of the in glass-forming alloys~\cite{Lebedev2017}.

In our previous work~\cite{Vasin2021}  we proposed a soft solidification model which can describe a glass state as a vector topologically degenerated field correspondent to the density of defects. This field is reinforced by the additional scalar phase field  correspondent to the crystalline ordering which in the first  approximation was written for the soft second-order transition. The combination of these two models reproduces an exponential time-scale relaxation of the defects and concurrence of the ``crystallization'' and vitrification fields. Moreover, this model was based on the  concept of glass represented as a frozen system of a topologically stable structure excitations (vortices)~\cite{Vasin2022}. However, the disordered phase relaxation was found to be an Arrhenius-type, which does not correspond to the experimentally observed relaxation behavior in the undercooled metallic liquids and glasses. 

Below we propose an extension of the model~\cite{Vasin2022}, where we introduce a crystallization kinetics with the first-order phase transition as a short-range ordering. The vitrification kinetics naturally emerges   in our simulations from topological vector field showing the competition of glass and crystalline  states during the quenching with finite cooling speeds.

\section{Concurrence of glass and crystal}

\subsection{Model of Glass transition}

In preset paper we consider a liquid within the model proposed by Yakov Frenkel~\cite{F}.  This approach is based on the postulate of the partial resemblance of crystals and liquids, and reveals many important properties. In the Frenkel's theory the liquid's particles at moderate temperatures assumed to behave in a manner similar to ones in a crystal phase. However, while in crystals atoms oscillate around their nodes, in liquids, after several periods, the particles change their positions.

The validity of this approach was proofed with the fact, that experimentally measured elastic properties of liquid are well distinguished on short spatio-temporal scales. In particular, in liquids a finite value of the shear modulus and a solid-like oscillation spectrum are observed at high frequencies~\cite{B5,B6,B7,B8,B9,B10,B11,B12}. Since transverse phonons can exist only in the liquid phase at frequencies exceeding the value of the inverse relaxation time (which decreases with the temperature growth), then the main distinctive criterion between a quasi-gas (soft) fluid and a  solid-like one (Frenkel liquid) is that the shear modulus for entire spectrum in soft fluid has zero value. On the phase diagram the region of the crossover from one liquid type to another is called Frenkel line \cite{B1,B2}.
According to this, the system can be considered as a solid if the shear modulus is different from zero in the entire frequency domain.
Thus, at low temperatures according to the Frenkel's approach, we consider a liquid as an elastic media containing both elastic and plastic deformations. The presence of the plastic deformations provides fluidity, and the elastic deformations defines the system's free energy.

The idea of formulation of two fields is close to the mosaic multistate scenario and  is more promisable comparing to the single phase description \cite{Cavagna2007}. So the introduction of the second  field describing the short range ordering $\varphi$ in addition to the topological vector field ${\bf A}$ can solve some of the crucial problems in the theoretical description of the melts  undergoing crystallization and vitrification.

The free energy density of a deformed elastic system is written as follows:
\begin{align}\label{I1}
\mathcal{F}=\dfr{\lambda}{2}u_{ll}^2+\mu \hat{\bf u}^2,
\end{align}
where $\lambda $ is the bulk modulus, $\mu$ is the instantaneous shear modulus, $\hat{\bf u}=u_{ij}={\mathrm{d} u_j}/{\mathrm{d} x_i}=\nabla_iu_j$ is the distortion tensor (${\bf u}$ is the strain vector), $u_{ll}$ is the designation of the diagonal components of $\hat{\bf u}$. The shear modulus $\mu $ is the microscopic parameter correspondent to the macroscopic shear modulus limit on long times. 
One can consider that shear modulus near glass transition is a linear function: $\mu_{eff}=\mu+\mbox{const}\cdot(T_0-T)$, where $T_0$ is some effective temperature parameter. This parameter can be read as  a temperature correspondent to the Frenkel line~\cite{B1,B2} at which the shear elasticity appears in the liquid~\cite{Vasin2022}.

The liquid here is considered as an elastic disordered media with presence of lots of moving plastic deformation cores correspondent to the dislocations and disclinations.  In the static the media is in mechanical equilibrium, however, the elastic energy of such system is non-zero since its disordered structure is geometrically frustrated and contains a number of stressed regions caused by local topologically protected distortions. The topologically protected rotation distortion corresponds to the disclination (or vortex line), and for simplicity below we consider only such distortion type. Besides, since the disclination is caused by a violation of axial symmetry, then, according to the topology, they are linear objects, and so the correspondent interaction field  is Abelian~\cite{Vasin2022}.

To proceed to the further formulation of the Hamiltonian let the system contain a disclination at the point ${\bf r}_n$. It breaks the simple connectivity of the space and leads to the distortion tensor which irreducible part corresponds to the rotation around considered disclination:
\begin{align}\label{I2}
\oint \hat {\bf u} \mathrm{d}{\bf l}=\int \nabla\times \hat {\bf u}\,\mathrm{d}^2{\bf r}={\bf \Omega}\delta_{{\bf r=r}_n}^{(2)},
\end{align}
where the space integration is performed over dimensionless variable ${\bf r}$: $V^{-1}\int\mathrm{d}V=\int \mathrm{d}^3{\bf r}$ for $|{\bf r}|<1$, and ${\bf \Omega}$ is the Frank vector. 

If the system contains $N$ vortices with a free energy density $\mathcal{F}$, \e{I1}, then its partition function can be represented as the functional integral of Hamiltonian $\mathcal{H}$:
\begin{align*}
W= \int \mathcal{D}\hat{\bf u}\exp\left[-\beta\int\mathrm{d}^3{\bf r}\,\mathcal{H}\right]\prod\limits_{n=1}^N\delta\left( {\bf l}\cdot\nabla\times\hat{\bf u}_{{\bf r}_n}-{\bf \Omega} J_{{\bf r}_n}\right),
\end{align*}
where $\beta=1/k_bT$, $\delta (\ldots )$ is the functional delta-function, ${\bf l}$ is a unit vector correspondent to the disclination director ${ J}_{{\bf r}_n}=0;\pm 1$.
Using the integral representation of the delta-function:
\begin{align}
W= \iint \mathcal{D}\hat{\bf u}\mathcal{D}{\bf A}\exp\left[-\beta\int\mathrm{d}^3{\bf r}\,\mathcal{H}\right].
\label{partitionfun}
\end{align}
where ${\bf A}$ is an auxiliary field, describing the set of defects, with condition defined by Eq.~(\ref{I2}). After the correspondent substitutions the effective Hamiltonian density of the vortex field takes the following form:
\begin{align}\label{A}
\mathcal{H}=\dfr12{\mu}{\hat{\bf u}}^2+
i\beta^{-1}{\bf A}\cdot\left[{\bf l}\cdot\nabla\times\hat{\bf u}-{\bf \Omega} \sum\limits_{n=1}^{N}J\delta_{{\bf r=r}_n}^{(2)}\right],
\end{align}
where $N$ is the quantity of the disclination elements, and ${\bf r}_n$ ($n=1$, 2,$\ldots $, $N$) are their coordinates~\cite{Vasin2022}.

\subsection{Coupling of the models}
To combine the topological ${\bf A}$-field   with the phase-field one can write Eq.~(\ref{partitionfun}) for full Hamiltonian $\mathcal{H}_{full}$:
\begin{align}\label{WHfull}
W= \iint \mathcal{D}\varphi\mathcal{D}\hat{\bf u}\mathcal{D}{\bf A}\exp\left[-\beta\int\mathrm{d}^3{\bf r}\,\mathcal{H}_{full}\right] , \quad \quad \textrm{where}
\end{align}
\begin{align}\label{ACr}
\mathcal{H}_{full}=\dfr12{\mu}{\hat{\bf u}}^2+
i\beta^{-1}{\bf A}\cdot\left[{\bf l}\cdot\nabla\times\hat{\bf u}-{\bf \Omega} \sum\limits_{n=1}^{N}J\delta_{{\bf r=r}_n}^{(2)}\right] \nonumber\\
+\dfr12\alpha{\hat{\bf u}}^2\varphi^2+(\nabla \varphi)^2+a(T-T_c)\varphi^2-b\varphi^4+c\varphi^6.
\end{align}
Here we introduce the linear part of short-range ordering ($\varphi$) linearly expanded around $\varphi=0$ in a row to describe the crystallization  in a manner proposed in mean-field theories~\cite{Ryzhov2017,bib:pe}. The effective shear modulus $\mu_{eff}=\mu +\alpha\varphi^2$ from Eq.~(\ref{ACr})  is the temperature dependent function linearly growing with temperature decreasing  since quadratic term $\varphi^2$ is proportional to   $ (T_c-T)$ in the mean-field models~\cite{Kostorz2001,Ryzhov2017,Alexander1978}.

\subsection{Simplification and assumptions}
After integration of \e{WHfull} over $\hat{\bf u}$-field of elastic distortions one can obtain 
\begin{align*}
W= \iint \mathcal{D}\varphi\mathcal{D}{\bf A}\prod\limits_{\bf r}\left(\sqrt{\dfr{2\pi\beta^{-1}}{\mu+\alpha\varphi_{\bf r}^2}}
\exp\left[-\beta\mathcal{H'}_{full}({\bf r})\right]\right),
\end{align*}
\begin{align}
\mathcal{H'}_{full}= \dfr{\beta^{-2}}{2(\mu+\alpha\varphi^2)}(\nabla\times {\bf A})^2-i\beta^{-1}{\bf \Omega} {\bf A}\sum\limits_{n=1}^{N}J\delta_{{\bf r=r}_n}^{(2)}\nonumber\\
+(\nabla \varphi)^2+a(T-T_c)\varphi^2-b\varphi^4+c\varphi^6 +\dfr1{2\beta}\ln \left(1+\mu^{-1}\alpha\varphi^2\right),
\label{Z}
\end{align}
which corresponds to the system of vortices that can interact by the ${\bf A}$-field. Here $\alpha$ is a coupling parameter and $\mu+\alpha$ is a crystalline (dense phase) elastic coefficient.
After integration of $\mathcal{H'}_{full}$ over ${\bf A}$-field for two vortices, one could find the interaction betwen the as the coulomb one, which is analogous to the interaction in our previous model~\cite{Vasin2021}.

We consider the arbitrary number of vortices, therefore one can average over the grand canonical ensemble. After this averaging the system effective Hamiltonian density assumes the form:
\begin{align}
\mathcal{H}^*_{full}= \dfr{\beta^{-2}}{2(\mu+\alpha\varphi^2)}(\nabla\times {\bf A})^2-g\beta^{-1}\cos\left({\bf \Omega  A}\right)\nonumber\\
+(\nabla \varphi)^2+a(T-T_c)\varphi^2-b\varphi^4+c\varphi^6\nonumber\\
+\dfr1{2\beta}\ln \left(1+\mu^{-1}\alpha\varphi^2\right).
\label{Z1}
\end{align}
where  $g$ is the vortices density, which is the Hamiltonian density in the sine-Gordon theory~\cite{I9,Minnhagen}.


Let one expand the cosine term into the power series of \e{Z1} for ${\bf A}$. In the quantum field theory for the $d$-dimensional system close to the critical point, the only Taylor expansion terms with powers less than $2d/(d-2)=6$ are relevant~\cite{Zee}. It means, that  for the three dimensional case $d=3$ one can only take into account first two terms and get rid of the higher ones. Thus the fluctuation corrections are relevant only for these first two terms, and thereby  the system effective Hamiltonian (\ref{Z1})  can be written as follows:
\begin{align}
&\mathcal{H}^*_{full}=\dfr{\beta^{-2}}{2(\mu+\alpha\varphi^2)}(\nabla\times {\bf A})^2\nonumber\\
&+g\beta^{-1}\left({\bf \Omega A} \right)^2\left(\dfr{1}{2}
-\dfr{\left({\bf \Omega A}\right)^2}{4!}+\dfr{\left({\bf \Omega A}\right)^4}{6!}\right) 
+\nonumber \\
&(\nabla \varphi)^2+a(T-T_c)\varphi^2-b\varphi^4+c\varphi^6
+\dfr1{2\beta}\ln \left(1+\mu^{-1}\alpha\varphi^2\right).
\end{align}
Then one can split ${\bf A}$-field into the  fast, $\tilde{\bf A}$, and the slow, ${\bf A}$, contributions: ${\bf A}\to {\bf A}+\tilde{\bf A}$, and average the whole domain over $\tilde{\bf A}$, whis will also lead to a constant value of $ {\bf \Omega}$.
Considering the average $\langle \tilde{\bf A}\rangle=0$ let one rewrite the above expression as follows:
\begin{align}\label{O1}
&\mathcal{H}^*_{full}=\dfr{\beta^{-2}}{2(\mu+\alpha\varphi^2)}(\nabla\times {\bf A})^2  \nonumber\\
&+g\beta^{-1}\dfr{\Omega^2}{2}{\bf A}^2  \left(1-\dfr{\Omega^2 }{2} \langle \tilde{\bf A}\tilde{\bf A}\rangle_0\right)\nonumber\\
&-g\beta^{-1}\dfr{\Omega^4}{4!}{\bf A}^4\left(1-\dfr{\Omega^2 }{2} \langle \tilde{\bf A}\tilde{\bf A}\rangle_0\right)+g\beta^{-1}\dfr{\Omega^6}{6!}{\bf A}^6 \nonumber \\
&+(\nabla \varphi)^2+a(T-T_c)\varphi^2-b\varphi^4+c\varphi^6
+\dfr1{2\beta}\ln \left(1+\mu^{-1}\alpha\varphi^2\right),
\end{align}
where
\begin{align}\label{O2}
\langle \tilde{\bf A}\tilde{\bf A}\rangle_0=\lambda_D\int\limits_0^{\lambda_D^{-3}}\dfr{\mathrm{d}^3{\bf q}}{(2\pi)^3}\dfr{(\mu+\alpha\varphi^2)\beta}{{\bf q}^2}\nonumber\\
=\int\limits_0^{1}\dfr{\mathrm{d}^3{\bf p}}{(2\pi)^3}\dfr{(\mu+\alpha\varphi^2)\beta}{{\bf p}^2}= \dfr{(\mu+\alpha\varphi^2)\beta}{2\pi^2}.
\end{align}
Here we define the Debye length, $\lambda_D$, and the dimensionless momentum, ${\bf p}=\lambda_D{\bf q}$. Considering $M^2 =gk_b(T-T^*)$ as the square of the effective ${\bf A}$ field ``mass'' or inertial response, one can formulate the final Hamiltonian of the system undergoing the phase transition to the solid phase with a concurrence of glassy and crystalline phases:
\begin{align}
&\mathcal{H}^*_{full}=\dfr{\beta^{-2}}{2(\mu+\alpha\varphi^2)}(\nabla\times {\bf A})^2 \nonumber\\
&+M^2\left(\dfr12(\Omega{\bf A})^2
- \dfr1{4!}(\Omega{\bf A})^4 \right) +g\beta^{-1}\dfr{\Omega^6}{6!}{\bf A}^6
\nonumber \\
&+(\nabla \varphi)^2+a(T-T_c)\varphi^2-b\varphi^4+c\varphi^6
+\dfr1{2\beta}\ln \left(1+\mu^{-1}\alpha\varphi^2\right).
\label{EfH}
\end{align}
The phase transition temperature in the disclination subsystem is related to the temperature of the vitrification $T^\ast$ and defined as
\begin{align}\label{Tg}
T^*=\left(\dfr{\Omega}{2\pi}\right)^2\left(\mu+\alpha\varphi^2\right).
\end{align}
To  study presented Hamiltonian \e{EfH}  we formulated dynamical equations and then performed a set of numerical simulations to show the concurrence between the formation of glassy and crystalline phases. Introduced in \e{EfH} physics  can be clarified in a short as: (i) variable $\varphi$ controls the short-range ordering (or dense packing), one have $\varphi=0$ for liquid and $\varphi=1$ for solid phase; (ii) variable ${\bf A}$ is a auxiliary topological field existing in dense phase defining the shear on long ranges, this field describes the presence of the defects as topological vortices and other inhomogeneities (for $|{\bf A}|\neq 0$).


\subsection{Dynamical equations}

\onecolumngrid

\begin{center}
\begin{table}
\centering
\caption{Model parameters utilized in the numerical simulations.\label{tab}}%
\begin{tabular*}{500pt}{@{\extracolsep\fill}lcccccccccccc@{\extracolsep\fill}}
\toprule
    		& $\alpha$  & $\mu$  & $\epsilon$  & $k_B$ & $a$ & $b$ & $c$  & $g$  & $\Omega$ & $\Gamma_{\varphi}$  & $\Gamma_A$   & $T_c$     \\
 Values	   &  1 & 1 	  & 1 	&  0.5		& 1		& 0.2				& 0.1					& 1		& 0.2	&	1				& 1				& 0.5   \\
\bottomrule
\end{tabular*}
\end{table}
\end{center}
\twocolumngrid

In the presence of the thermal fluctuations, the unconserved kinetics is described in terms of the stochastic non-equilibrium dynamics \cite{Hohenberg1977,PP,Vasilev1998}.
The following kinetic equations can be derived from the variational derivative of the full Hamiltonian \e{EfH} with the addition of the stochastic sources dependent on $T$:
\begin{align}
\label{mov1}
&-\Gamma_{\varphi}\partial_t \varphi=\dfr{\delta \mathcal{H}^*_{full}}{\delta \varphi }=-\dfr{\alpha\beta^{-2}\varphi}{\left(\mu+\alpha\varphi^2\right)^2}(\nabla{\bf A})^2 \nonumber \\
&-2gk_b\alpha\left(\dfr{\Omega}{2\pi}\right)^2\varphi\left(\dfr12(\Omega{\bf A})^2
- \dfr1{4!}(\Omega{\bf A})^4 \right)-\nabla^2 \varphi \nonumber \\
&+2a(T-T_c)\varphi-4b\varphi^3+6c\varphi^5
+\dfr{\alpha\varphi}{\beta\left(\mu+\alpha\varphi^2\right)} + k_b T N_\varphi ,\\
\label{mov2}
&-\Gamma_{A}\partial_t {\bf A}=\dfr{\delta \mathcal{H}^*_{full}}{\delta {\bf A}}=-\dfr{\beta^{-2}}{2(\mu+\alpha\varphi^2)}\nabla^2{\bf A}\nonumber \\
&+M^2\left(\Omega{\bf A}- \dfr1{3!}(\Omega{\bf A})^3 \right) +g\beta^{-1}\dfr{\Omega^6}{5!}{\bf A}^5 + k_b T N_A,
\end{align}
where $N_\phi$ and $N_A$ are the noise generating variables with normally distributed value probability and amplitudes equal to 1. The thermal fluctuations are delta-correlated, so one can note $\langle N_j N_j\rangle=\delta(t)\delta({\bf r})$ for $j=[\varphi; {\bf A}]$.
We presume the parameters of the Eqs.~(\ref{mov1})-(\ref{mov2}) to be as shown in Table~\ref{tab}.

The moving equations Eqs.~(\ref{mov1})-(\ref{mov2}) include the driving force control parameter  $T$. The temperature here not only provide the certain driving force pushing the system to the equilibrium following the Lyapunov condition~\cite{gj19}, but also induces  the phase nucleation with the fluctuation mechanism. Meanwhile the fluctuations in this model play a significant role during freezing, while them determine the inhomogeneities  relaxation speed. We will demonstrate it below in the Results section.

\subsection{Numerical implementation}
\label{25}
We performed the numerical simulations of the dynamical equations Eqs.~(\ref{mov1})-(\ref{mov2}) which were obtained using the Hamiltonian \e{EfH}  in the two-dimensional computational domain with periodic boundary conditions. The initial conditions were set as a constant distributions for $\varphi=0$ and random distribution for $|{\bf A}|= 2$. This exact initial distribution corresponds to the disordered liquid phase full of randomly distributed defects.
The noise sources were obtained from the random function generator with the uniform distribution and zero average values. The computational domain consisted of $L=100\times 100$ dimensionless units with up to $50$ grid points along the edges; the maximum triangle mesh element size was set as $\ell=2$. We tested the mesh convergence and for the specific cooling speeds there were no significant influence on the size of the topological peculiarities during solidification process. The domain size was also sufficient to proceed to the formation of the bulk multigrain phase.
 
To compute given problem we utilized the COMSOL Multiphysics 6.0 software~\cite{comsol6} with MUMPS direct solver and the backward differentiation formula for time integration. The time step were set as $\delta t=1.5\cdot10^{-6}$, this value was find consistent for the dissipation of the noise  contribution and  enough for the convergence. To take into account the acceptable range of the variables we introduced segregated solver with fixed limits on $\varphi$ which controlled the adaptive time step for certain iterations.  All calculations were performed on two-processor AMD Epyc-based computer.

\section{Results and discussion}

\subsection{Crystallization}

\begin{figure}[h]
  \centerline{
\includegraphics[width=\columnwidth]{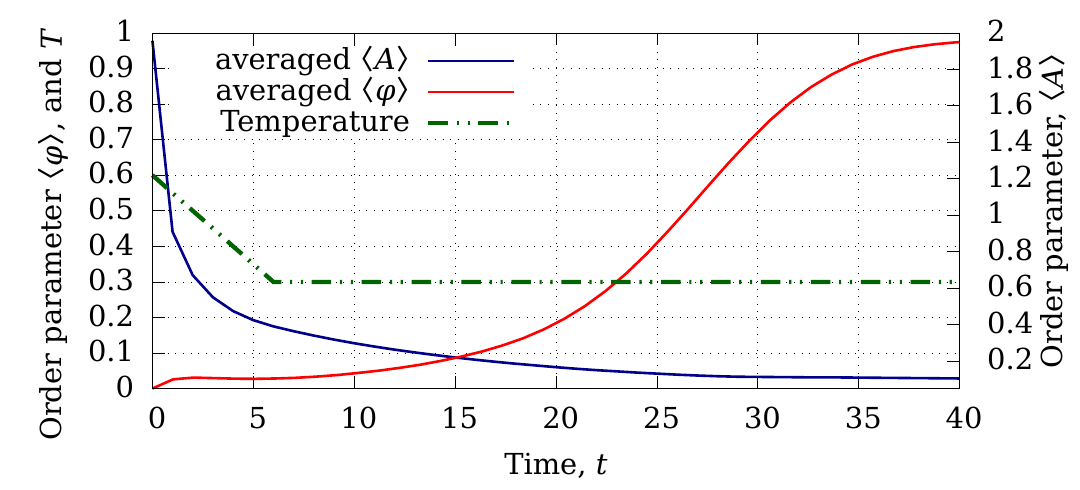} }
  \caption{\label{nucl} Dependence of the  averaged  order parameters $\langle A \rangle$,   $\langle \varphi \rangle$ and temperature $T$  on $t$  during the crystallization; the structural relaxation during the extremely rapid quenching is presented; $T_s=0.6$; $T_f=0.3$;  $V_{cool}=5\cdot 10^{-2}$.}
\end{figure}

\onecolumngrid

\begin{figure}[!h]
  \centerline{
  \resizebox{.99\columnwidth}{!}{%
  (a)\includegraphics[height=5cm]{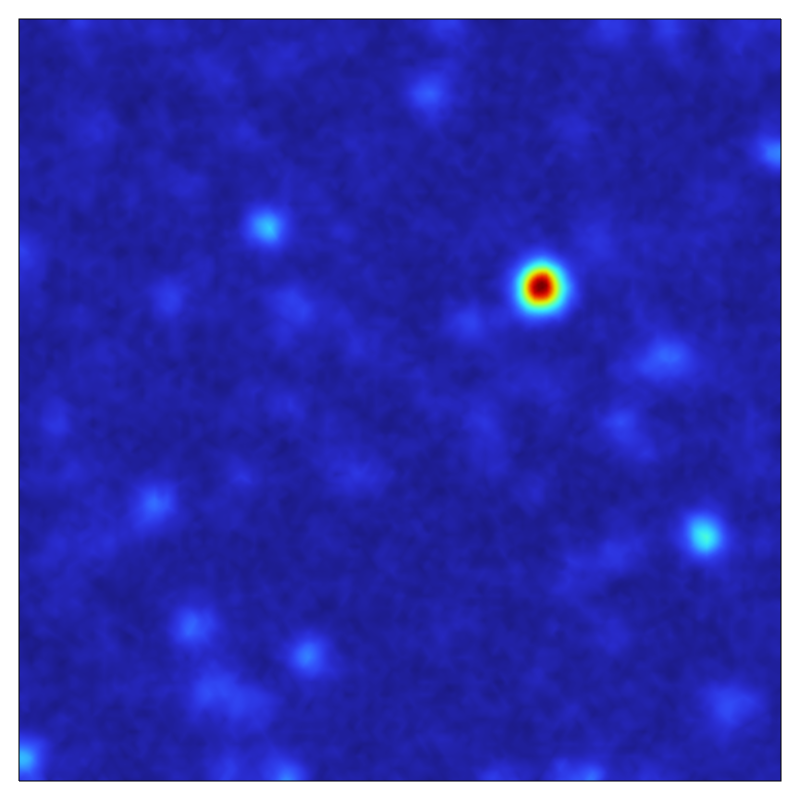}
  (b)\includegraphics[height=5cm]{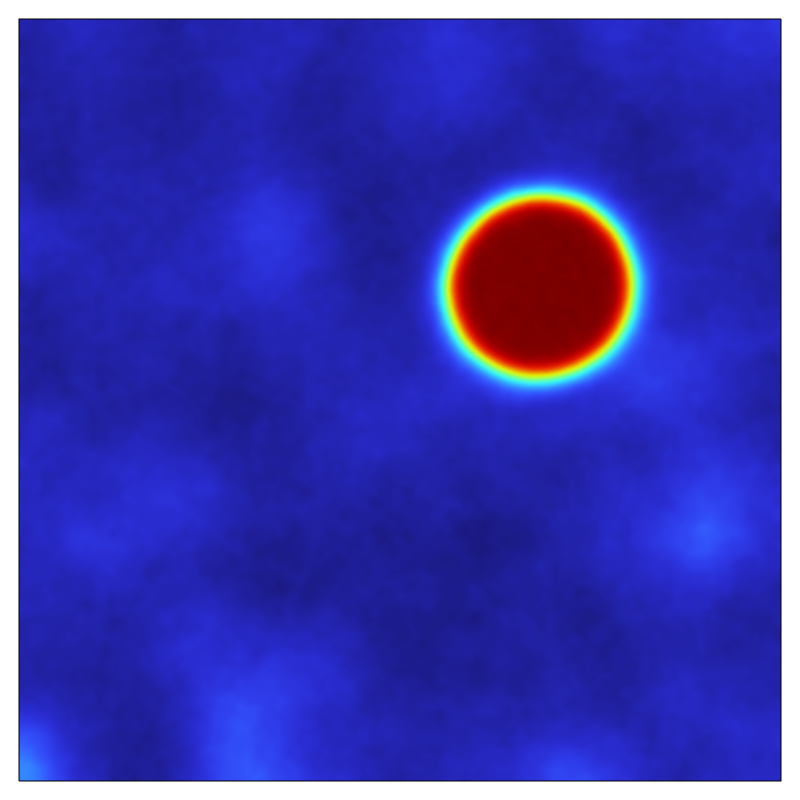}
  (c)\includegraphics[height=5cm]{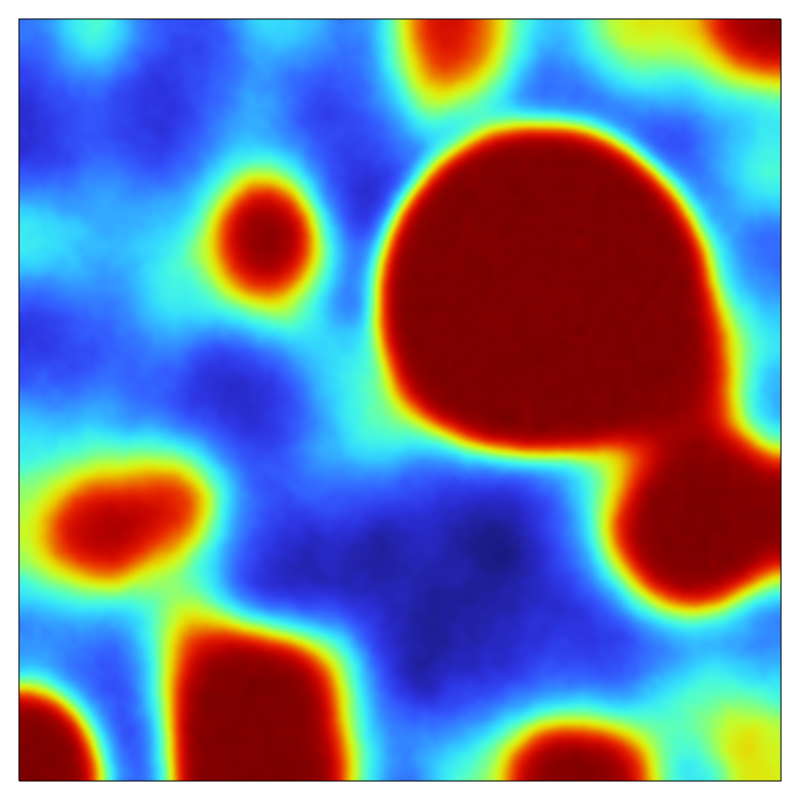}
  (d)\includegraphics[height=5cm]{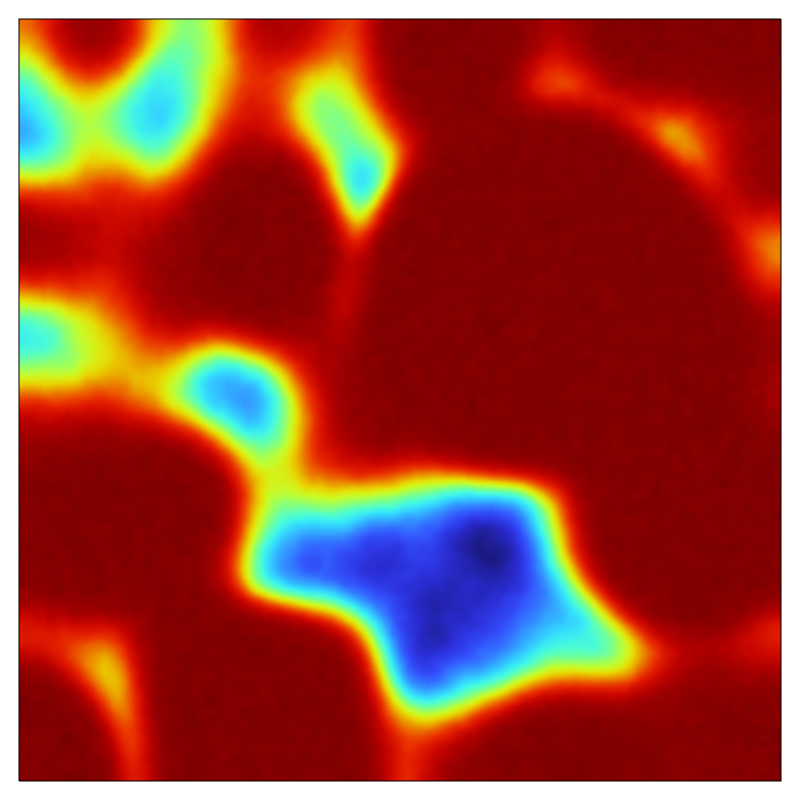}
  \includegraphics[height=5cm]{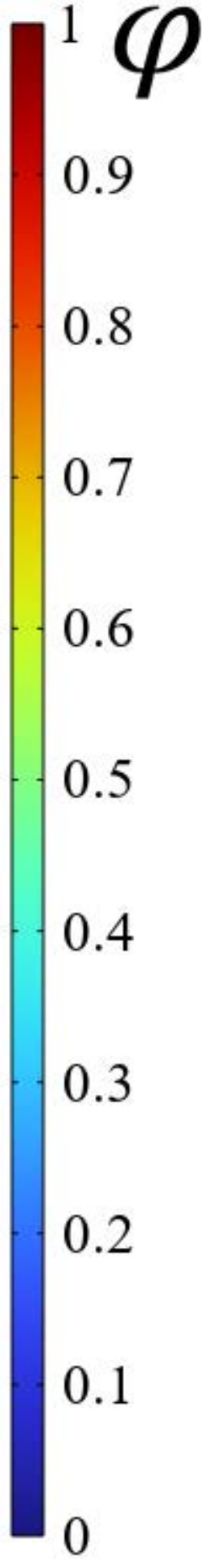}
  }}
  \caption{\label{nucls} Snapshots of the order parameter $\varphi$ during the homogeneous  nucleation of crystalline phase; $T_f=0.6$; $T_f=0.3$; $V_{cool}=5\cdot 10^{-2}$; (a) $t=3$; (b) $t=20$; (c) $t=28$; (d) $t=30$. This snapshots correspond to the Fig.~\ref{nucl}.}
\end{figure}
\twocolumngrid

We simulated a solidification to the crystalline phase by gradually cooling at the finite speed $V_{cool}$. 
The simulation started from high temperatures $T=T_s>T_c$  with initial disordered liquid phase in a bulk, then we set temperature $T$ as a function of  $V_{cool}$, initial $T_s$ and finishing $T_f$ temperatures as
  \begin{equation}
T=  \begin{cases}
      T_f + V_{cool}(t_f-t),& \quad \textrm{if} \quad 0\leqslant t<t_f;\\
      T_f, & \quad \text{otherwise}.
    \end{cases}
\end{equation}
Here $t_f$ is the cooling end time, $t_f=(T_f-T_s)/V_{cool}$. Let one define undercooling as
\begin{equation}
\Delta T = T_c - T,
\end{equation}
where the critical temperature $T_c$ is related to the first order phase transition described by $\varphi$-field. 

Presented undercoolings $\Delta T$ (driving forces) should be understood as a closely related to the realistic undercoolings obtainable in the laboratory experiments. Besides, the key contradiction to the essence of the experimental cooling  aligns with the difference in the temperature behavior during the quenching. In laboratory quenching the speed of local cooling is limited by the heat transport coefficients of the media and / or cooling surface. Thus the cooling intensity is determined by the heat flux. In our model we are trying to implement this mechanism with the assumption of  linear decreasing of $T$ with finite speed $V_{cool}$. Yet we can not fully describe the mechanism of the quenching with this approach, since there is no joint heat transfer problem solved, and then one can not interpret $V_{cool}$ as a general cooling speed observable in the experiments.  The laboratory cooling speed is determined by the heat flux, here we only define a temperature drop rate. So in our computational problem with $T$ change we define a global driving forces in micro-volume considering the infinite the speed of heat propagation.

It is also important to note that in our model the phase change during  the slow cooling occurs gradually  undergoing sequential transitions between  profitable states. In the realistic alloy melt starts to ``freeze'' below the glass transition temperature  due to the drastic decrease of the mobility of atoms \cite{Galenko2019rsta,Galenko2019a}.
The static properties of the specific Hamiltonian \e{EfH} demonstrate that the first order transition barrier is controlled by the coefficient $b$ and can be found as  $T_c+\frac{b^2}{4ac}$ which comes from the definition of the melting temperature.  The consequence of having such barrier can be seen in Fig.~\ref{nucl} as a slow response of the $\varphi$-subsystem on the applied driving force caused by $\Delta T$. The vector field of defects reacts much faster, however, we are not going to look in detail at competition in boundary dynamics of $\varphi$ and ${\bf A}$ as soon as we studied that earlier on a very close model~\cite{Vasin2021}. The interesting demonstration of the homogeneous nucleation in the undercooled melt can be found in Fig.~\ref{nucls}. Here one can find an exponential nucleation driven by the fluctuation mechanism as for the pure first-order phase transition. 

For the given set of parameters the thermodynamically preferred phase is the crystalline one $\varphi=1$, nevertheless on very long $t$ the possible occurrence of the equilibrium phase at ${\bf A}=const$ is possible. However the physical meaning of ${\bf A}$ becomes debatable here and it requires additional study of thermodynamically consistent values for the competitive phases in the region of small $\Delta T$ and unambiguous crystallization region. For $T<T^\ast$ one gets a grain structure or an amorphous phase which will be shown in the next section.

\subsection{Vitrification}

\begin{figure}[!h]
  \centerline{
\includegraphics[width=\columnwidth]{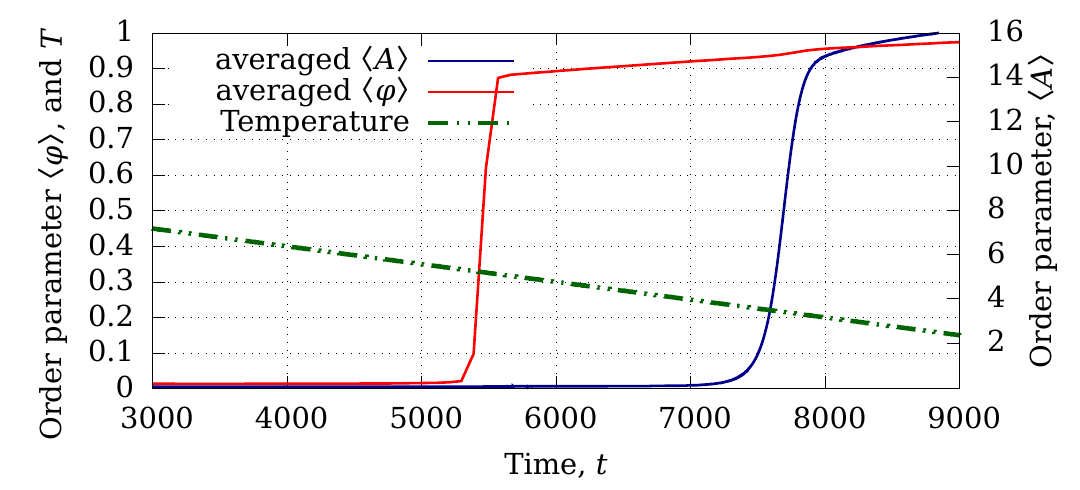} }
  \caption{\label{vitr} Dependence of the averaged  order parameters $\langle A \rangle$,   $\langle \varphi \rangle$ and temperature $T$  for the  slow solidification on $t$; the consequent  ordering and late growth of ${\bf A}$-field with the formation of the coarse grained  polycrystalline phase (see Fig.~\ref{vitrs}); $T_s=0.6$; $T_f=0.1$; $V_{cool}=0.5\cdot 10^{-4}$.}
\end{figure}

\onecolumngrid

\begin{figure}
  \centerline{
  \resizebox{.99\columnwidth}{!}{%
  (a)\includegraphics[height=5cm]{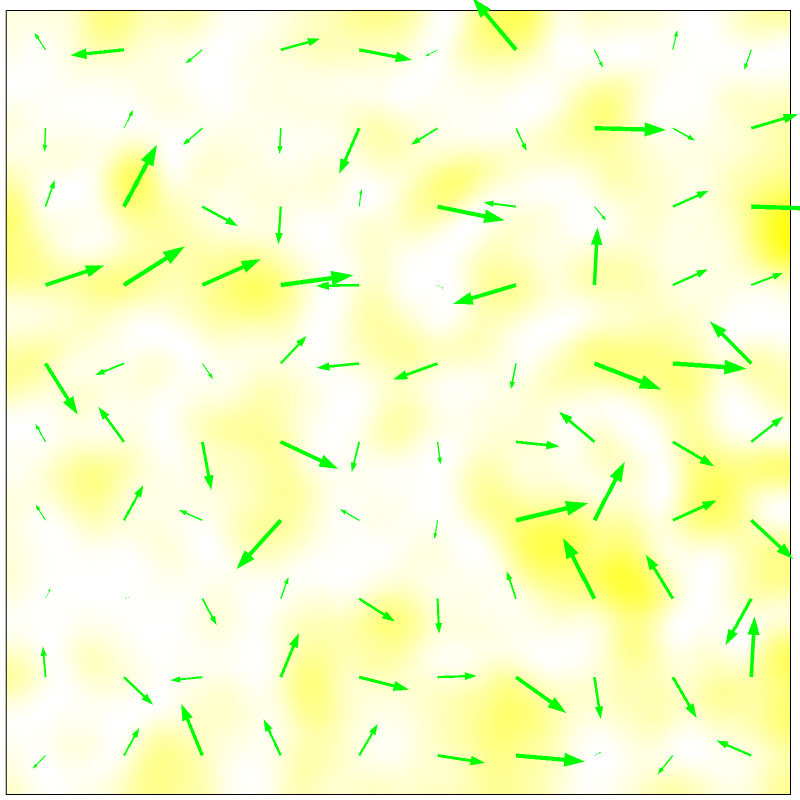}
  (b)\includegraphics[height=5cm]{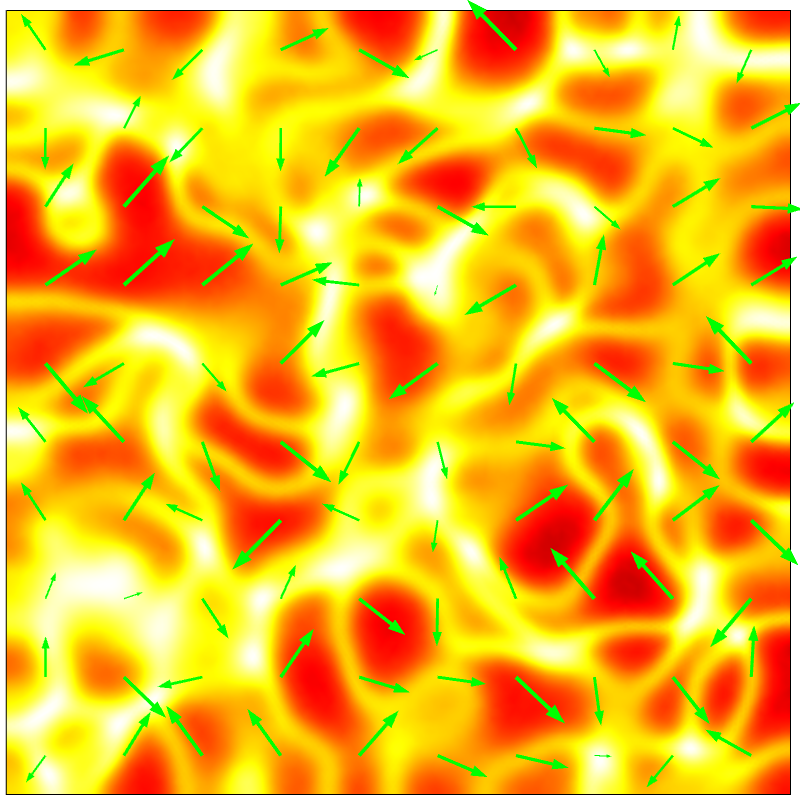}
  (c)\includegraphics[height=5cm]{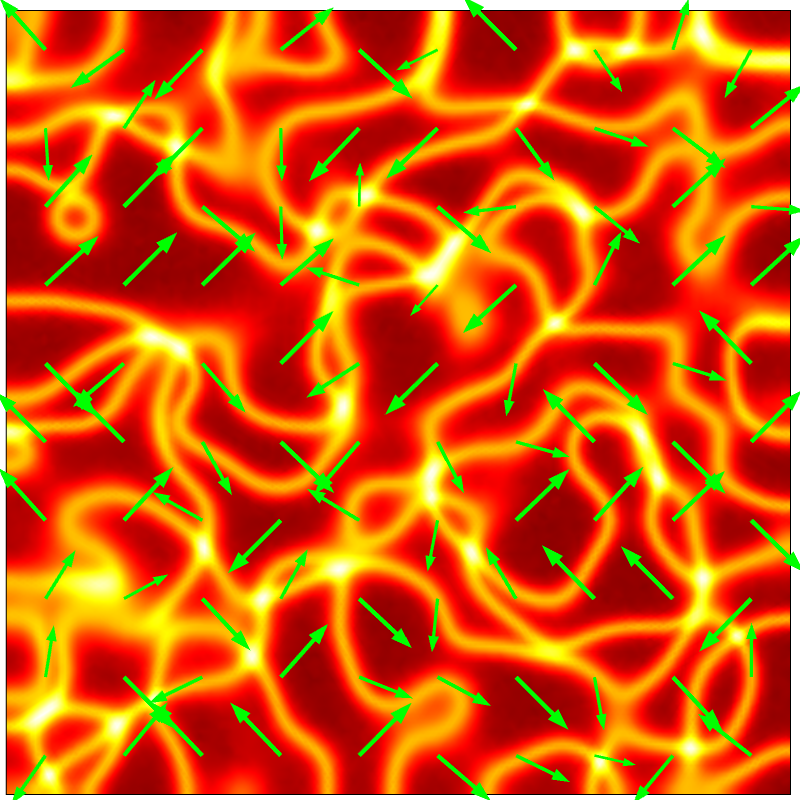}
  (d)\includegraphics[height=5cm]{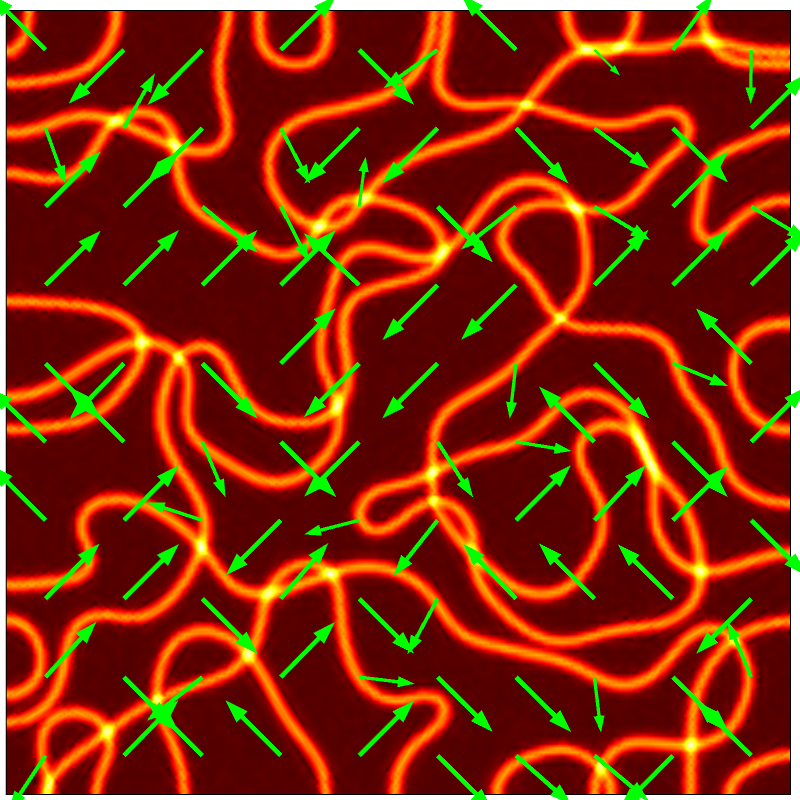}
  \includegraphics[height=5cm]{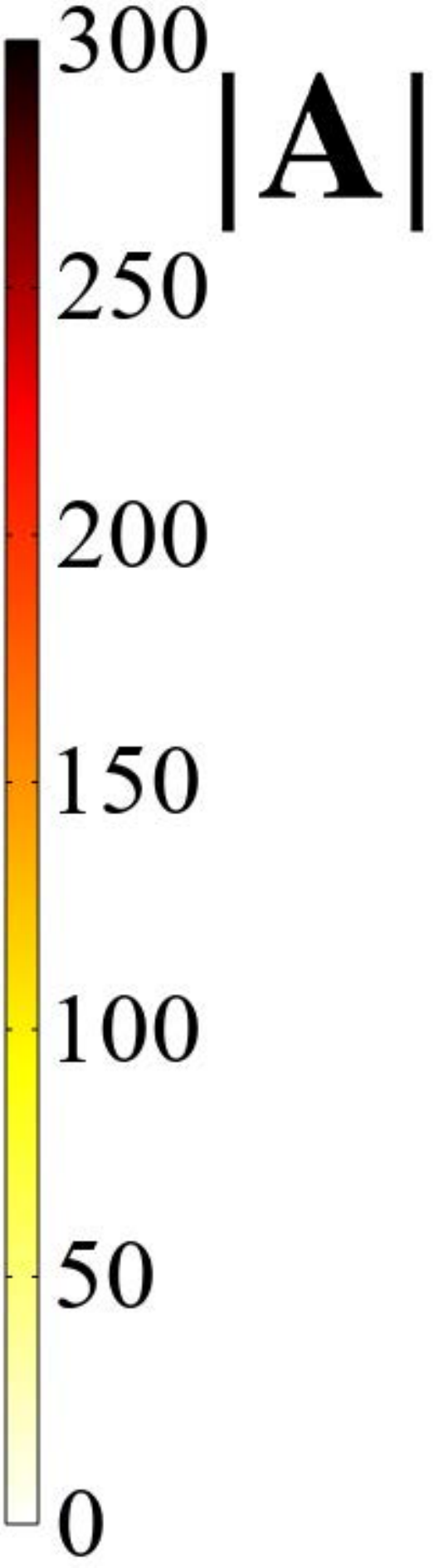}
  }}
  \caption{\label{vitrs} Snapshots of the module of field $|{\bf A}|$ during  slow quenching resulting the formation of coarse grained polycrystalline structure, green arrows correspond to the x- and y-components of the vector ${\bf A}$; $T_s=0.6$; $T_f=0.1$; $V_{cool} = 0.5\cdot 10^{-4}$; (a) $t=7600$; (b) $t=7750$; (c) $t=7850$; (d) $t=8500$. This snapshots correspond to the Fig.~\ref{vitr}.}
\end{figure}
\twocolumngrid

  To study the different scenarios of the vitrification during quenching at different $V_{cool}$ we fixed the initial and final temperatures as $T_s = 0.6$, $T_f  =0.1$. Here $T_f<T^\ast$  [\e{Tg}], so the conditions are favorable for the start of the vitrification when the correspondent regimes are reached. In Fig.~\ref{vitr} one can find the results of the simulation of  slow quenching, which lead to the coarse grained crystalline structure [Fig.~\ref{vitrs}(d)]. One can find that there is enough time for the grain growth, where grains are determined by the continuous regions of the equally oriented ${\bf A}$ separated by the curved borders. This grain boundaries can be read as a regions of presence of inhomogeneities and defects which is typical for this kind of aggregations. The surface energy is controlled by the $\frac{\beta^{-2}}{2(\mu+\alpha\varphi^2)}$ term from \e{mov2} correspondent to the elastic energy and thus surface tension. There is a tendency to coarsening of the dispersed structures, so one can observe a slow grain growth, which is nevertheless limited due to the different grain's orientation. The intense growth of the averaged module of the topological defects field $|{\bf A}|$ starts right after reaching the temperature of  glass transition $T^\ast=0.274$.

  \begin{figure}
  \centerline{
  \resizebox{\columnwidth}{!}{%
  (a)\includegraphics[height=5cm]{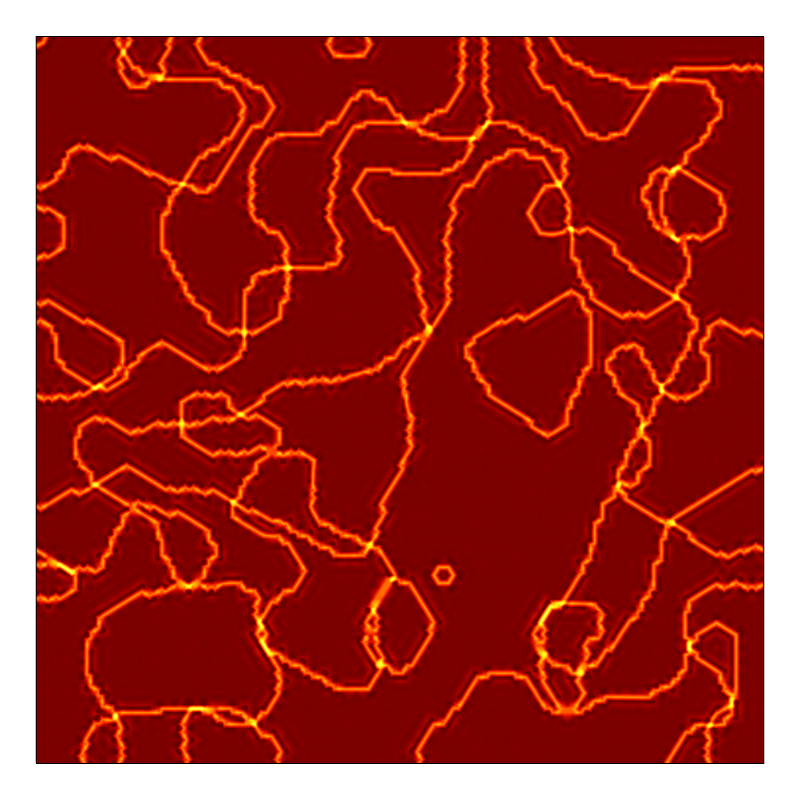}
  (b)\includegraphics[height=5cm]{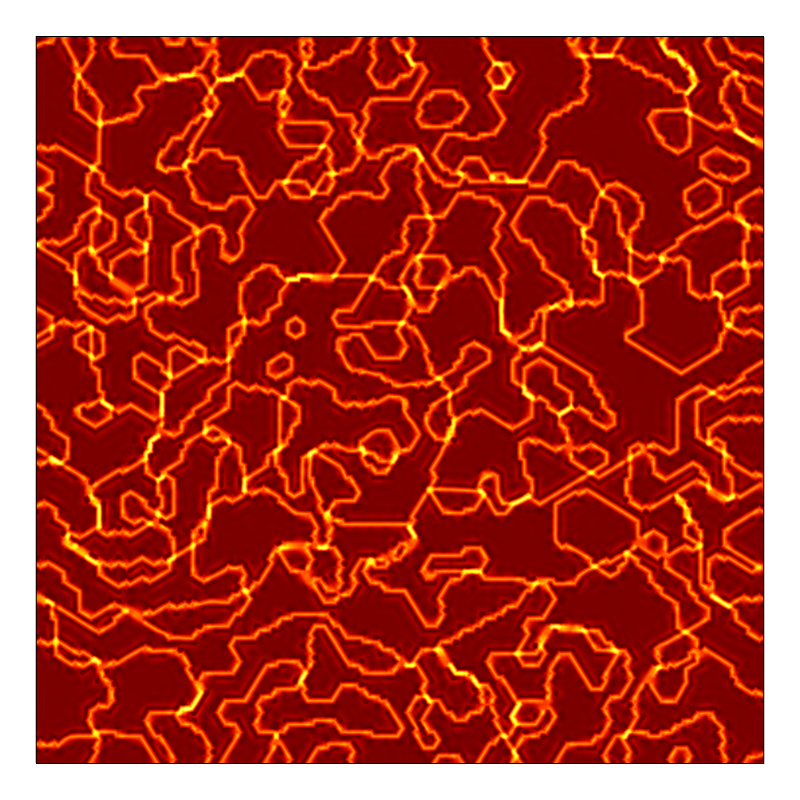}
  (c)\includegraphics[height=5cm]{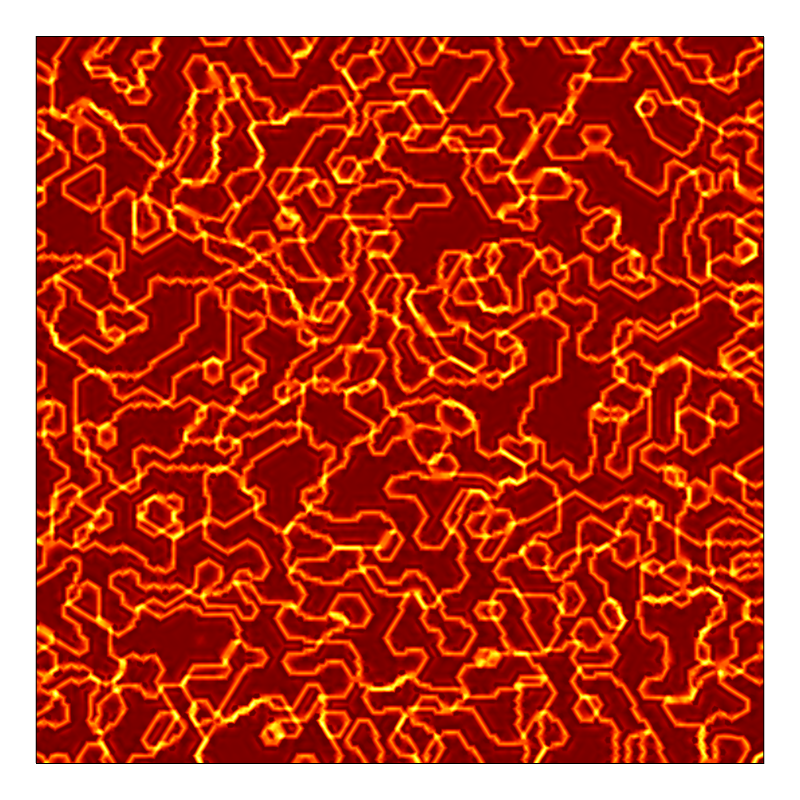}
  \includegraphics[height=5cm]{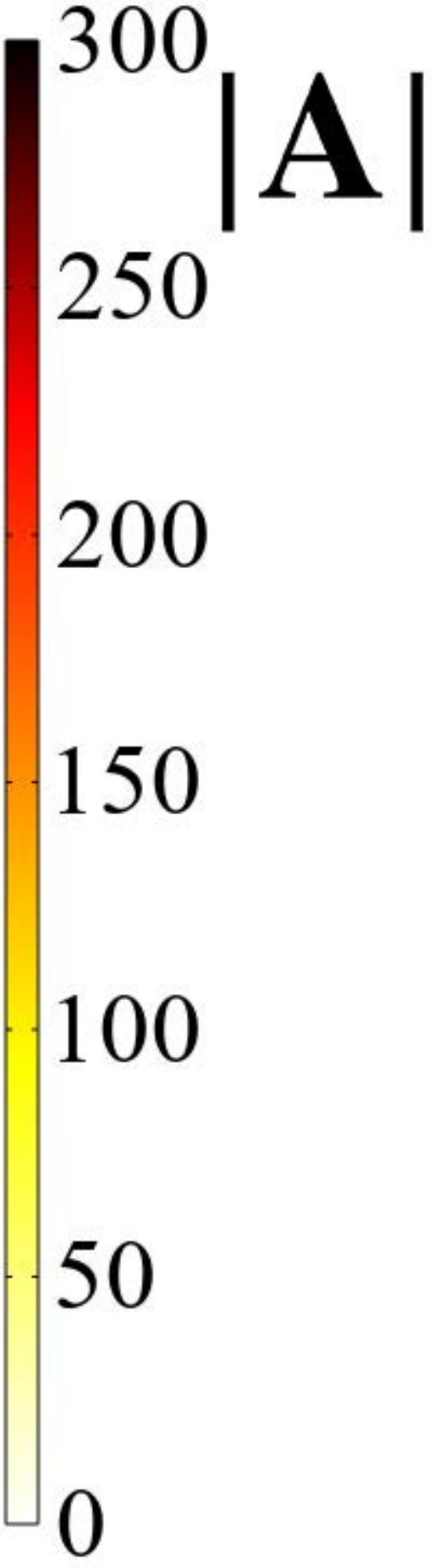}
  }}
  \caption{\label{dcool3} Snapshots of the module of $|{\bf A}|$ during the  quenching at different speeds $V_{cool}$ presenting  the formation of polycrystalline and amorphous structure; $T_s=0.6$; $T_f=0.1$; $t=20000$; (a) $V_{cool} = 1\cdot 10^{-4}$; (b) $V_{cool} = 5\cdot 10^{-4}$; (c) $V_{cool} = 15\cdot 10^{-4}$.}
\end{figure}

\begin{figure}
  \centerline{
\includegraphics[width=\columnwidth]{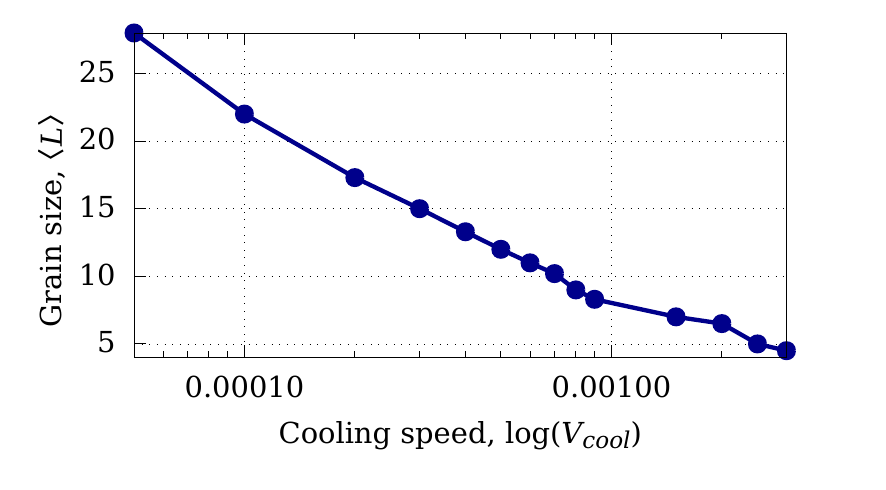} }
  \caption{\label{dcool} Dependence of the averaged  grain size $\langle L \rangle$ on the cooling speed $V_{cool}$ plotted in logarithmic scale; $T_s=0.6$; $T_f=0.1$.}
\end{figure}
  
Let one consider different $V_{cool}$, so one can find the correspondent averaged grain size $L$ which is to be a non-linear function. During the rapid quenching to the $T^\ast$ the formed grains start to coarse with a certain finite speed. However, when $T$  drops below $T^\ast$ the fluctuation-driven ${\bf A}$ rotations from the adjacent grains become less active (see Fig.~\ref{dcool3}). Further temperature decreasing permanently freezes structure preserving the  grain structure. The fine grain structure demonstrates the metastable solid media full of the space-stretched defects which in fact is the amorphous structure or glass.
 One can calculate the average grain size $\langle L \rangle$ as  $V_{cool}$ function (see Fig.~\ref{dcool}). Such exponential curve is more or less common for the known fraction of the amorphous and crystalline phases in rapidly solidifying melts~\cite{Kharanzhevskiy2022}. Meanwhile, we found an important uncertainty based on the continuous formulation of the model. As soon as the finest grain size is limited by the finite element size, one can not definitively  formulate lower limit of $\langle L \rangle$ (as a function of $V_{cool}$). In principle it could be limited by the relation of mobility coefficients $\Gamma_A$ to the elastic constants, and thus surface tension. It is important to note, in the presented  model the nonlinear response on the time-dependent effect of $V_{cool}$ is implemented only with a dynamical coupling of two dissipative fields. The fluctuations here naturally results to some sort of memory function~\cite{book}. 
  
\section{Conclusions}

In the present work, we demonstrated the concurrence of the vitrification and formation of the crystalline phase during the rapid quenching using the coupled phase-field and topological glass transition model. Presented approach, due to the introduction of the fluctuations, allows one to show a complex behavior and nonlinear kinetic response to the rapidly changing control parameter (temperature $T$). 
The formulation of the model allows description of the first-order phase transitions and nucleation. The model qualitatively describes the key features of the kinetics of the order-disorder states competition, we found the temperature of glass transition and proved a significant impact of the cooling speed on the regimes of structure formation. The numerical simulations resulted that the solidification during quenching leads to the sequential selection of the dense phase (solidification, $T<T_c$), then the disordered phase (vitrification, $T<T^\ast$).  In the case of low undercoolings and low cooling speed this process leads to the coarsening of the fine grain structure and formation of the polycrystalline phase; the fine grained structure with uniformly distributed defects formed at large $V_{cool}$ could  be interpreted as a glassy phase. Presented results agree with the kinetics of known glass-forming alloys.

The further developing of the current model should clarify  the several important questions such as dependence of the glass transition temperature on the cooling speed, which in the presented framework could be possible to achieve within the implementation of the temperature dependent mobility (diffusion) coefficients. Another interesting problem is the development of the model supporting the concentration diffusion which may potentially introduce non-Arrhenius relaxation behavior in the undercooled binary metallic liquids and glasses due to the additional degree of freedom correspondent to the solute redistribution.


\section*{Acknowledgments}
The work was supported by the RnD project "Artificial Intelligence
in the development, training and maintenance of expert systems for knowledge use in natural, technical sciences and humanities"  AAAA-A19-119092690104-4 of Udmurt Federal Research Center, Ural Branch of the Russian Academy of Science; the computational part was supported by the Council of the President of the Russian Federation for State Support of Young Scientists (Grant No. MD-6103.2021.1.2).

\end{document}